\documentclass[twocolumn,prl,preprintnumbers,superscriptaddress,amsmath,amssymb,english]{revtex4}
\usepackage{amsmath,amssymb,amsfonts,mathrsfs}
\usepackage{amsthm}
\usepackage{CJK}
\usepackage{bm}
\usepackage{babel}
\usepackage{graphicx}
\usepackage{makecell}
\allowdisplaybreaks
\usepackage{braket}
\usepackage[breaklinks]{hyperref}
\usepackage{booktabs}
\usepackage{multirow}
\usepackage{float}
\usepackage{gensymb}
\usepackage{rotating}
\hypersetup{colorlinks=true, linkcolor=blue, citecolor=blue, filecolor=blue, urlcolor=blue}
\UseRawInputEncoding

\begin{document}

\title{Magnetic field induced Valley-Polarized Quantum Anomalous Hall Effects in Ferromagnetic van der Waals Heterostructures }

\author{Fangyang Zhan}
\affiliation{Institute for Structure and Function $\&$ Department of Physics, Chongqing University, Chongqing 400044, P. R. China}
\affiliation{Chongqing Key Laboratory for Strongly Coupled Physics, Chongqing 400044, P. R. China}

\author{Baobing Zheng}
\email{scu$_$zheng@163.com}
\affiliation{College of Physics and Optoelectronic Technology, Baoji University of Arts and Sciences, Baoji 721016, P. R. China}
\affiliation{Institute for Structure and Function $\&$ Department of Physics, Chongqing University, Chongqing 400044, P. R. China}
\affiliation{Chongqing Key Laboratory for Strongly Coupled Physics, Chongqing 400044, P. R. China}

\author{Xiaoliang Xiao}
\affiliation{Institute for Structure and Function $\&$ Department of Physics, Chongqing University, Chongqing 400044, P. R. China}
\affiliation{Chongqing Key Laboratory for Strongly Coupled Physics, Chongqing 400044, P. R. China}

\author{Jing Fan}
\affiliation{Center for Computational Science and Engineering, Southern University of Science and Technology, Shenzhen 518055, P. R. China}

\author{Xiaozhi Wu}
\affiliation{Institute for Structure and Function $\&$ Department of Physics, Chongqing University, Chongqing 400044, P. R. China}
\affiliation{Chongqing Key Laboratory for Strongly Coupled Physics, Chongqing 400044, P. R. China}

\author{Rui Wang}
\email{rcwang@cqu.edu.cn}
\affiliation{Institute for Structure and Function $\&$ Department of Physics, Chongqing University, Chongqing 400044, P. R. China}
\affiliation{Chongqing Key Laboratory for Strongly Coupled Physics, Chongqing 400044, P. R. China}
\affiliation{Center for Quantum materials and devices, Chongqing University, Chongqing 400044, P. R. China}
\affiliation{Center for Computational Science and Engineering, Southern University of Science and Technology, Shenzhen 518055, P. R. China}

\begin{abstract}
The valley-polarized quantum anomalous Hall effect (VQAHE) attracts intensive interest since it uniquely combines valleytronics and spintronics with nontrivial band topology.
Here, based on first-principles calculations and Wannier-function-based tight-binding (WFTB) model, we reveal that valley-based Hall effects and especially the VQAHE induced by external magnetic fields can occur in two-dimensional (2D) ferromagnetic van der Waals heterostructures (vdWHs). The results show that considerable valley-splitting derived from the Zeeman exchange energy
drives these vdWHs generating the valley anomalous Hall effect and then the VQAHE. The chiral edge states and quantized Hall conductance are utilized to confirm the presence of VQAHE. Besides, it is also found that external electric fields (or tuning interlayer distances) can facilitate the realization of VQAHE, and thus we present a phase diagram in a broad parameter regime of magnetic fields and electric fields (or interlayer distances). Our work not only offers a class of ferromagnetic vdWHs to realize various valley-based Hall phases, but also can guide advancements for designing topological devices with spin-valley filtering effects based on the VQAHE.
\end{abstract}



\maketitle
The extrema on the valence/conductuion band, known as valleys, provide a different degree of freedom to encode and manipulate information analogous to charge and spin, thus stimulating extensive interest in so-called valleytronics both theoretically and experimentally \cite{Xu2014,Schaibley2016,Small2018}. Especially, the experimental discovery of two-dimensional (2D) graphene and transition metal dichalcogenides (TMDs) significantly promotes advancements of valleytronics. These 2D valleytronic materials possess readily accessible valley degrees of freedom since their intervalley interactions are nearly absent.
Because graphene crystallizes in an inversion ($\mathcal{I}$) symmetric honeycomb lattice,  the occurrence of its valley-dependent phenomena, such as the valley-dependent optical selection rule and valley Hall effects (VHE), requires to break the $\mathcal{I}$-symmetry \cite{PhysRevLett.99.236809,PhysRevB.77.235406}. This always brings great challenges for the realization of valleytronics in graphene-based systems despite many efforts devoting to overcome these difficulties \cite{2009APL,PhysRevLett.106.136806,PhysRevLett.110.046601,PhysRevLett.106.156801,vila2021valleypolarized}.
Alternatively, the ideal materials to realize valley physics are TMDs due to its intrinsic $\mathcal{I}$-symmetry breaking. Therefore, the valley polarization, which is the prerequisite for valleytronics, can more easily achieved in TMDs \cite{PhysRevLett.108.196802,Cao2012}.

\begin{figure}
\setlength{\belowcaptionskip}{-0.30cm}
    \centering
    \includegraphics[width=8.6cm]{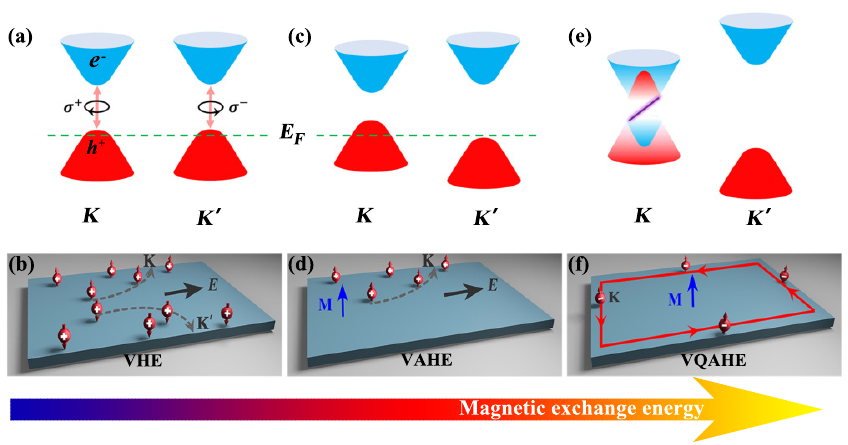}
    \caption{Schematic illustration for different responses to the $K$ and $K'$ valleys under a magnetic field. In (a) and (b), the hole pockets at two valleys spilled over the Fermi level can be driven by a longitudinal in-plane electric field, forming the valley Hall effect (VHE). In (c) and (d), the Zeeman exchange energy lifts the valley degeneracy in a ferromagnetic system with intrinsic magnetic moment $\mathbf{M}$, leading to the valley-polarized anomalous Hall effect (VAHE). In (e) and (f), a stronger magnetic field can trigger band inversion only at the $K$ valley, giving rise to the VAQHE with one valley-dependent chiral edge state.
\label{figure1}}
\end{figure}
An intriguing and important valley-based phenomenon is the VHE \cite{Gorbachev448,Mak1489}, of which the $K$ and $K^{'}$ valleys of hexagonal Brillouin zone host opposite Hall currents for hole/electron carriers, as shown in Figs. \ref{figure1}(a) and \ref{figure1}(b). The VHE originates from the fact that the valley-contrasting Berry curvature acts as an effective magnetic field in momentum space. As a result, carriers move in opposite directions under an in-plane electric field, giving rise to transverse Hall currents. As is well known, the time-reversal ($\mathcal{T}$) symmetry is closely related to various Hall effects. For example, a $\mathcal{T}$-invariant 2D topological insulator possess topologically protected helical edge states, known as the quantum spin Hall effect \cite{PhysRevLett.95.226801,PhysRevLett.96.106802}, and breaking the $\mathcal{T}$-symmetry of 2D topological insulator with magnetic doping results in the quantum anomalous Hall effect \cite{PhysRevLett.101.146802,Chang167}. Combining valleytronics and band topology, a more promising phenomenon, i.e., the valley-polarized quantum anomalous Hall effect (VQAHE), can be present \cite{PhysRevLett.106.156801,PhysRevB.91.045404,PhysRevB.97.085420,PhysRevLett.119.046403,PhysRevLett.127.116402}. The key issue to realize the VQAHE is to destroy the $\mathcal{T}$-symmetry for valleytronic materials. Based on this principle, to date, many proposals have been explored, such as introducing external magnetic fields \cite{PhysRevLett.112.106802,PhysRevB.91.045404,PhysRevB.92.155419,PhysRevB.101.155425}, decorating magnetic transition metal atoms \cite{PhysRevLett.119.046403}, achieving ferromagnetic order by half-hydrogenating honeycomb lattice \cite{PhysRevB.91.165430,PhysRevB.91.041303}, and intrinsic ferromagnetism \cite{doi:10.1021/acs.chemmater.1c00798}. Among these approaches, the application of external magnetic fields is a direct way to generate valley-splitting due to the Zeeman effect. However, Zeeman exchange energy is often rather small \cite{PhysRevLett.113.266804,2014Magnetic}, and thus relatively weak valley-splitting limits the formation of VQAHE.

Recently, magnetic proximity effect in van der Waals heterostructures (vdWHs) stacked with TMDs and 2D ferromagnets has been revealed as an effective and feasible approach to generate large valley-splitting \cite{PhysRevB.92.121403,PhysRevB.101.205404,PhysRevB.101.125401,2018nano}. In particular, with the progress of growth technology for 2D intrinsic ferromagnetic semiconductors, such as atomic layers of CrI$_3$ \cite{Huang2017} and  Cr$_2$Ge$_2$Te$_6$ \cite{Gong2017}, unprecedented opportunity for studying the marriage of valleystronics and spintronics in ferromagnetic vdWHs is expected \cite{PhysRevB.100.085128,PhysRevB.100.195307}. While recent advancements have been very encouraging, the VQAHE originated from magnetic proximity effects has been very limited. Therefore, the exploration of an effective and controllable approach to achieve the VQAHE is highly desirable.

Here, we propose another scheme to realize the VQAHE. Under applying an external magnetic field to the ferromagnetic vdWHs, it is found that the Zeeman exchange energy with the aid of magnetic proximity effects can lead to considerable valley-splitting. As shown in Figs. \ref{figure1}(c) and \ref{figure1}(d), we can see that the Zeeman exchange energy lifts the valley degeneracy and induces different responses to the $K$ and $K'$ valleys in a ferromagnetic system with intrinsic magnetic moment $\mathbf{M}$,  consequently leading to the VAHE \cite{C9NR03315G}. Furthermore, with increasing Zeeman energy of  magnetic fields, the valley-splitting makes the bands around the $K$ valley undergoing a band inversion process and then evolving into the nontrivial regime, while the bands around the $K'$ valley always preserve the trivial feature. As a result, the VQAHE with one valley-dependent chiral edge states can be present [see Figs. \ref{figure1}(e) and \ref{figure1}(f)].

\begin{figure}
\setlength{\belowcaptionskip}{-0.30cm}
    \centering
    \includegraphics[width=8.6cm]{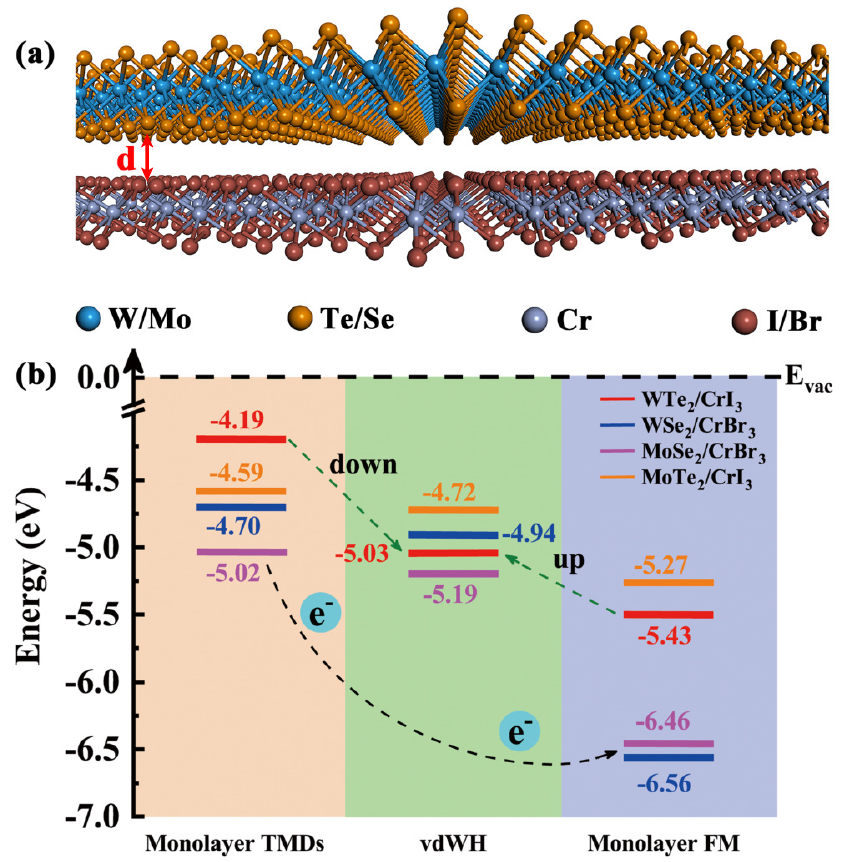}
    \caption{(a) Schematic of the MX$_{2}$/CrY$_{3}$ vdWHs with a interlayer distance \textit{d}. (b) The directions of charges transfer and the shift of the Fermi level in the process of forming heterojunctions. The Fermi levels are marked respectively as red, blue, purple and orange line in WTe$_{2}$/CrI$_{3}$, WSe$_{2}$/CrBr$_{3}$, MoSe$_{2}$/CrBr$_{3}$ and MoTe$_{2}$/CrI$_{3}$ vdWHs, respectively.
    \label{figure2}}
\end{figure}

First, we construct the ferromagnetic vdWHs by depositing the monolayer 2H-phase TMDs MX$_2$ (M = Mo/W; X = Se/Te) on the recently reported 2D CrY$_3$ (Y = I, Br), 
as illustrated in Fig. \ref{figure2}(a). To elucidate electronic and topological properties of these ferromagnetic vdWHs, we employ first-principles calculations based on the density-functional theory \cite{PhysRev.140.A1133} [see details in Supplemental Material (SM) \cite{SM}]. 
In the main text, we use the WSe$_{2}$/CrBr$_{3}$ vdWH as an example to depict phase transitions of valley-based Hall effects induced by external magnetic fileds. The results of other vdWHs are given in the SM \cite{SM}. The CrY$_{3}$ has a hexagonal crystal structure in which per magnetic Cr ion with ${3\mu_B}$ magnetic moment.
Three representative stacking configurations between 2H-TMDs and CrY$_{3}$ substrates are considered (the detailed structures in Fig. S1 \cite{SM}). The most thermodynamically stable structures are considered in the following. The maximum lattice mismatch between the monolayer 2H-TMDs and CrY$_{3}$ substrates is only 2.5\%, thus 
facilitating their experimental fabrication.

\begin{figure*}
\setlength{\belowcaptionskip}{-0.30cm}
\setlength{\abovecaptionskip}{-0.00cm}
    \centering
    \includegraphics[width=18cm]{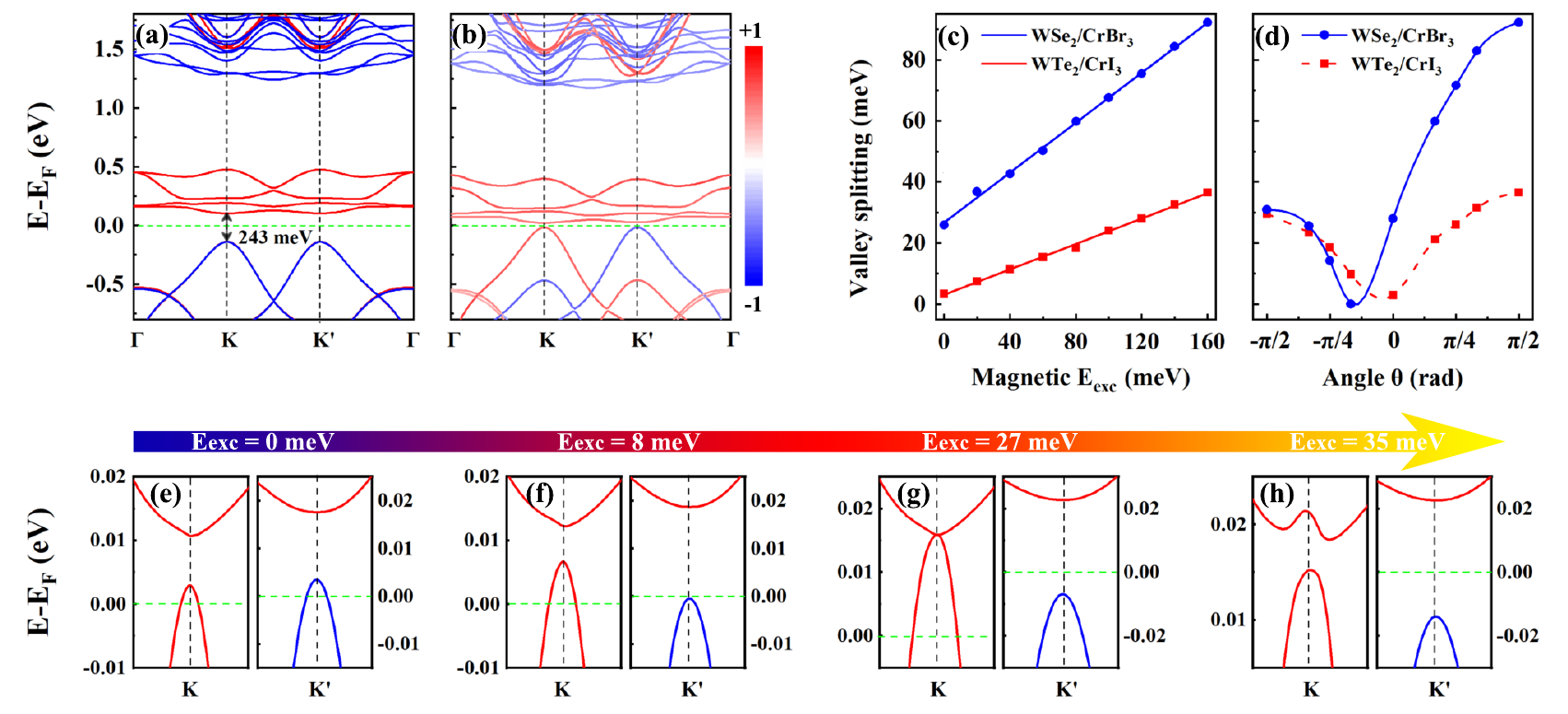}
    \caption{The spin-resolved band structures of WSe$_{2}$/CrBr$_{3}$ vdWH (a) without  and (b) with  SOC in the absence of external magnetic fields. The valley-splitting as functions of (c) the magnitude and (d) the direction of magnetic field. (e-h) The enlarged views of band structures evolving with the Zeeman exchange energy ($\rm E_{exc}$) around the $K$ and $K^\prime$ valleys. Red (blue) lines denote spin-up (spin-down) channels. Here, a perpendicular electric field is fixed at 0.1 (V/\AA).
   \label{figure3}}
\end{figure*}

The stability and charge transfer of these 2D vdWHs can be evaluated by the binding energy and the work function, respectively. To assess the stability of these ferromagnetic 2D vdWHs, we calculate their binding energies, defined as  $E_{b}=E_{\rm vdWHs}-E_{\rm TMDs}-E_{\rm FM}$, where $E_{\rm vdWHs}$, $E_{\rm TMDs}$, and $E_{\rm FM}$ are total energies of vdWHs, monolayer 2H-phase MX$_2$,  and monolayer CrY$_3$, respectively. The results of $E_{b}$ (see Table S1 \cite{SM}) suggest that all the considered 2D vdWHs are experimentally feasible. To clarify the charge transfer between the 2H-phase MX$_2$ and CrY$_3$ layers, we calculate the work functions $W$ of these vdWHs, monolayer 2H-phase MX$_2$, and monolayer CrY$_3$ (The electrostatic potential of vdWHs is shown in Fig. S2 \cite{SM}). 
As shown in Fig. \ref{figure2}(b), the calculated results demonstrate that electrons in the 2H-TMDs layer can spontaneously flow into the adjacent CrY$_{3}$ layer due to the lower work functions of 2H-TMDs than those of CrY$_{3}$, thus leading to a vertical built-in electric field between two layers. Note that the previous study indicated that the vertical electric field combining with a magnetic field can significantly increase the valley-splitting in silicene \cite{PhysRevLett.109.055502,PhysRevB.90.035142}. Therefore, our proposed 2D ferromagnetic vdWHs subject to magnetic fields are expected to further enhance the valley-splitting due to cooperative effects of their intrinsic ferromagnetism and electric fields.

Next, we mainly focus on the WSe$_{2}$/CrBr$_{3}$ vdWH as a typical example to explore the valley polarization. As shown in Figs. \ref{figure3}(a) and \ref{figure3}(b), the band structures without and with spin-orbit coupling (SOC) of WSe$_{2}$/CrBr$_{3}$ vdWH are depicted, respectively. In the absence of SOC, the WSe$_{2}$/CrBr$_{3}$ vdWH exhibits semiconducting features with a moderate band gap of 243 meV. As SOC is included, the spin degeneracy is removed due to the broken spin-rotation symmetry. We can see that the large spin-splitting only occurs at the valence bands, which is in accordance with the monolayer TMDs \cite{PhysRevLett.108.196802}.
Moreover, the spin-projected band structures exhibit that two valleys possess opposite spin channels of valence bands near the Fermi level [see Fig. \ref{figure3}(b)]. That is to say, the spin is coupled to the valley degree of freedom, forming the so-called spin-valley lock effect \cite{Zeng2012,Mak2012}. 
Furthermore, an inspection of orbital contribution indicates hole and electron pockets at two valleys are respectively dominated by the $d_{xy}$ and $d_{x^{2}-{y^{2}}}$ orbitals of M atoms in the upper MX$_2$ layer and $d$ orbitals of Cr atoms in the lower CrY$_{3}$ layer (see Fig. S3 \cite{SM}), exhibiting type-\uppercase\expandafter{\romannumeral2} band alignment \cite{PhysRevB.87.075451}.

In order to elaborate the valley and topological properties under external magnetic fields, we construct the Wannier-function-based tight-binding (WFTB) model based on maximally localized Wannier functions (WFs) as implemented in the Wannier90 \cite{Mostofi2014}.
The WFTB Hamiltonian with SOC $H_{\rm SOC} $ is expressed as \cite{PhysRevB.98.075123,PhysRevB.101.035105}
\begin{equation}
\begin{split}
H_{\rm SOC,\alpha\beta}(\mathbf{k}) &= \big< \psi_{\mathbf{k},\alpha} | H_{\rm SOC} | \psi_{\mathbf{k},\beta} \big> \\
&= \sum_{\mathbf{R}} e^{i \mathbf{k} \cdot \mathbf{R}} t_{\alpha\beta} (\mathbf{R} - 0), \\
t_{\alpha\beta} (\mathbf{R} - 0) &= \big< \mathbf{0} + \mathbf{s}_{\alpha} | H_{\rm SOC} | \mathbf{R} + \mathbf{s}_{\beta} \big>.
\end{split}
\end{equation}
Here, $| \psi_{\mathbf{k}} \big>$ are Bloch states over the $\mathbf{k}$ space, $\mathbf{R}$ is the Bravais lattice vector, and $t_{\alpha\beta} (\mathbf{R} - 0)$ are the hopping parameters from orbital $\beta$ at site $\mathbf{s}_{\beta}$ in the home cell at $\mathbf{R}=0$ to orbital $\alpha$ at site $\mathbf{s}_{\alpha}$ in the unit cell located at $\mathbf{R}$. Then, we consider the magnetic field as a Zeeman term, and the Zeeman exchange energy is
\begin{equation}
H_{\rm Z} = g \mu_B (\mathbf{L}+2\mathbf{S}) \cdot \mathbf{B},
\end{equation}
in which $g$, $\mu_B$, $\mathbf{L}$, and $\mathbf{S}$ are the effective $g$-factor, Bohr magneton, orbital angular momentum, and spin angular momentum, respectively. Note that we here ignore the Peierls phase in the hopping parameters $t_{\alpha\beta}$ or Landau levels. This approach is reliable and effective for studying the magnetic field induced topological states \cite{PhysRevB.98.075123,PhysRevB.101.035105}. 

The valley-splitting is a crucial parameter for expediently utilizing the valley to realize the information processing, especially for the large valley-splitting \cite{Norden2019}. Thus, we next study the valley-splitting as functions of the magnitude [see Fig. \ref{figure3}(c)] and the direction [see Fig. \ref{figure3}(d)] of magnetic field. The valley-splitting $\Delta_{K,K^\prime}^{C,V}$ is defined as $\Delta_{K,K^\prime}^{C,V}=\lvert{\Delta_{K}^{C,V}-\Delta_{K^\prime}^{C,V}}\rvert$, where $\Delta_{K}^{C,V}$ and $\Delta_{K^\prime}^{C,V}$ are the exciton transition energies of the $K$ and $K^\prime$ valleys, respectively. As shown in Fig. \ref{figure3}(c), when the magnetic field is perpendicular to the plane of vdWHs, the valley-splitting increases linearly with the increase of Zeeman exchange energy, which is consistent with the experimental results \cite{2014Magnetic,PhysRevLett.114.037401}.
Moreover, the valley-splitting of WSe$_{2}$/CrBr$_{3}$ is more susceptible than that of WTe$_{2}$/CrI$_{3}$ under the magnetic field. Figure \ref{figure3}(d) shows the valley-splitting as a function of the direction of magnetic field for a given Zeeman exchange energy. Here, the direction of magnetic field is defined by the directional cosines $\vec{n}=(0, \cos \theta, \sin \theta)$, in which $\theta$ denotes the angle between the directional vector and the plane of the vdWHs. Because of magnetic proximity effects of the vdWHs, the valley-splitting exhibits an asymmetrical shape with the variation of the direction of magnetic field.
Therefore, the tunable valley-splitting induced by the Zeeman exchange energy with the aid of magnetic proximity effects provides a feasible way to manipulate the valley degree of freedom.

\begin{figure}
\setlength{\belowcaptionskip}{-0.30cm}
\setlength{\abovecaptionskip}{-0.00cm}
    \centering
    \includegraphics[width=8.6cm]{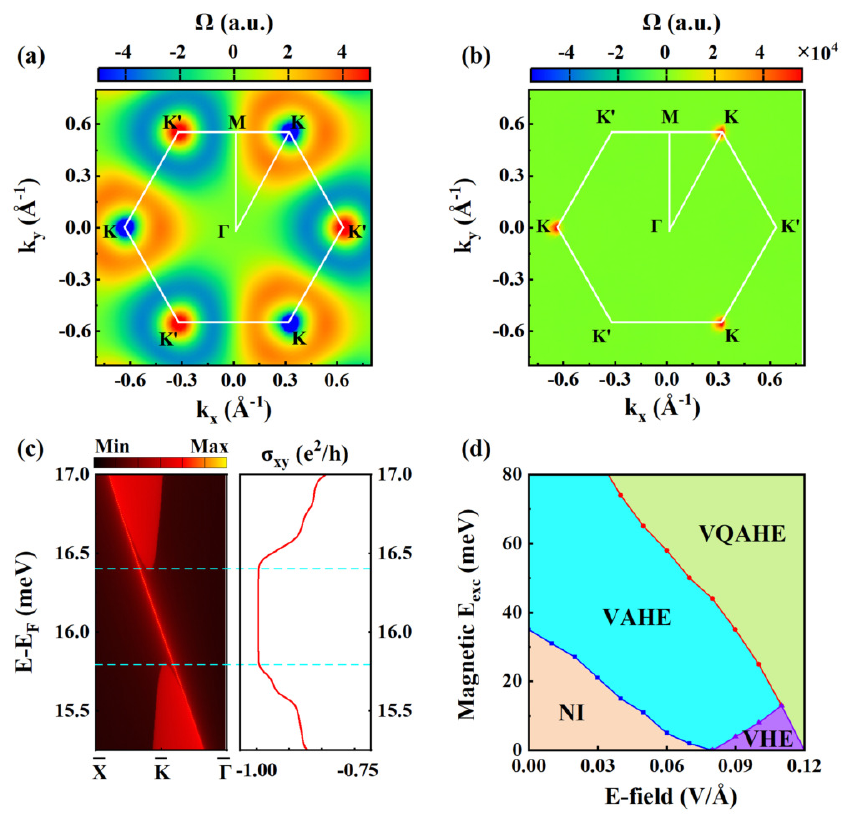}
    \caption{The Berry curvature distribution of WSe$_{2}$/CrBr$_{3}$ vdWH in the first Brillouin zone (a) without and (b) with an external magnetic field. In the panel (b), the Zeeman exchange energy is set to 35 meV and the VQAHE is present. (c) The semi-infinite LDOS of WSe$_{2}$/CrBr$_{3}$ vdWH with an external magnetic field, visibly exhibiting the chiral edge state, and the Hall conductance inside the nontrivial band gap exactly quantized to $-\textit e^{2}/h$. (d) The phase diagram for valley-based Hall effects of WSe$_{2}$/CrBr$_{3}$ vdWH as functions of the Zeeman exchange energy and electric field strength.
   \label{figure4}}
\end{figure}

Figures \ref{figure3}(e)-\ref{figure3}(h) show the evolution of band structure of WSe$_{2}$/CrBr$_{3}$ vdWH with increasing the Zeeman exchange energy. In the absence of external magnetic fields, the magnetic proximity effects lift the Kramers degeneracy of two inequivalent valleys, and the bands crosses the Fermi level both at the $K$ and $K^\prime$ valleys [see Fig. \ref{figure3}(e)], indicating the hole doping. An important hallmark to distinguish between valleys for these vdWHs without inversion symmetry is valley-contrasting Berry curvature, which can be calculated as \cite{PhysRevLett.49.405,PhysRevLett.92.037204,PhysRevB.74.195118}:
\begin{align}
\Omega(\mathbf{k}) &= \sum_{\mathbf{n}} f_{n} \Omega_{n}(\mathbf{k}), \\
\Omega_{n}(\mathbf{k}) &= -2\text{Im} \sum_{m \neq n} \frac{\hbar^{2} \big< \psi_{n\mathbf{k}} | v_{x} | \psi_{m\mathbf{k}} \big> \big< \psi_{m\mathbf{k}} | v_{y} | \psi_{n\mathbf{k}} \big>}{(E_{m}-E_{n})^{2}},
\end{align}
where the summation is over all of the occupied states, $f_{n}$ is the Fermi-Dirac distribution function, $E_{n}$ is the eigenvalue of the Bloch functions $\psi_{n\mathbf{k}}$, and $v_{x(y)}$ are the velocity operators. As shown in Fig. \ref{figure4}(a), the two valleys of WSe$_{2}$/CrBr$_{3}$ vdWH host nonzero and opposite Berry curvature $\Omega(\mathbf{k})$, which shows excellent consistency with previous studies \cite{PhysRevB.101.205404,PhysRevB.101.125401}. Therefore, spin-up holes from the $K$ valley and spin-down holes from the $K^\prime$ valley can be driven by a longitudinal in-plane electric field (\textbf{E}), inducing valley-contrasting transversal Hall currents perpendicular to \textbf{E} and forming the VHE [see Fig. \ref{figure1}(d)].

After an external magnetic field perpendicular to the plane of  WSe$_{2}$/CrBr$_{3}$ vdWH is applied, the Zeeman exchange energy leads to opposite response to the spin-up and spin-down bands. As a consequence, the valence bands around $K$ and $K^\prime$ valleys respectively move upward and downward. As the Zeeman exchange energy of magnetic field reaches to 8 meV, the bands near the $K^\prime$ valley presents the insulating behaviour, while the bands near the $K$ valley still preserves the metallic feature [shown in Fig. \ref{figure3}(f)]. In this case, only spin-up holes near the $K$ valley can produce transversal Hall currents, giving rise to the VAHE. With further increasing the Zeeman exchange energy, one can find that the bands at the $K$ valley first closes and then reopens, while the band gap at the $K^\prime$ valley is further enlarged [see Figs. \ref{figure3}(g) and \ref{figure3}(h)]; that is, the magnetic field triggers the topological phase transition from the VAHE to VQAHE. The critical gapless point at the $K$ valley occurs at the Zeeman exchange energy of 27 meV [see Fig. \ref{figure3}(g)]. This can be easily realized in experiments because of the large $g$ factor of WSe$_{2}$ \cite{2015Single}. We also calculate the evolution of Wannier charge centers based on the Wilson loop method \cite{PhysRevB.84.075119}, as show in Fig. S5 in the SM \cite{SM}, to verify the presence of this topological phase transitions. Similar topological phase transitions and the QAHE are also found in WTe$_{2}$/CrI$_{3}$, MoTe$_{2}$/CrI$_{3}$ and MoSe$_{2}$/CrBr$_{3}$ heterostructures (see the SM \cite{SM}), demonstrating that it is a generic strategy utilizing magnetic fields with cooperative magnetic proximity effects to realize the VQAHE in 2D ferromagnetic vdWHs.


To further confirm that such system can indeed realize the VQAHE, we calculate the Berry curvature of the WSe$_{2}$/CrBr$_{3}$ vdWH.
As depicted in Fig. \ref{figure4}(b), it is found that the Berry curvature around the $K$ valley diverges while that around the $K^\prime$ valley vanishes, which implies that the electrons flow in the curl field and contribute to the quantized Hall conductance. The Chern number $\mathcal{C}$, obtained by integrating the Berry curvature over the occupied states in the whole Brillouin zone (BZ), is  $\mathcal{C}=-1$. It is worth noting that the Chern number $\mathcal{C}$ is only from the $K$ valley, i.e., $\mathcal{C}_K=-1$ and $\mathcal{C}_{K^\prime}=0$, leading to the valley Chern number $\mathcal{C}_v=\mathcal{C}_K-\mathcal{C}_{K^\prime}=-1$ and thus suggesting the emergence of VQAHE.
The nonzero valley Chern number $\mathcal{C}_v$ of VQAHE corresponds to the topologically nontrivial edge states and the quantized Hall conductivity. Therefore, based on the WFTB Hamiltonian, we next employ the iterative Greens function method \cite{Sancho_1985} as implemented in WannierTools package \cite{WU2017} to calculate the semi-infinite local density of states (LDOS). 
As shown in the left panel Fig. \ref{figure4}(c), it can be seen that the topological chiral edge states connecting the valence and conduction bands are clearly visible. The anomalous Hall conductance $\sigma_{xy}$ can be computed by the Kubo formula using WFTB Hamiltonian \cite{Mostofi2014}. The right panel of Fig. \ref{figure4}(c) shows the $\sigma_{xy}$ exhibits exactly a quantized plateau of $-\textit e^{2}/h$ inside the nontrivial band gap.

The magnetic proximity effects can be tuned by external electric fields and interlayer distance \cite{PhysRevB.100.085128}. Therefore, to fully understand cooperative effects of magnetic fields and electric fields (or interlayer distance) on valley-based Hall effects and topological phase transitions, we calculate a phase diagram as functions of the Zeeman exchange energy and electric field strength, as shown in Fig. \ref{figure4}(d).
It is worth noting that the perpendicular electric field can effectively reduce the critical value of Zeeman exchange energy for realizing the VQAHE, thus facilitating to achieve the valley-splitting induced spin- and valley-polarized dissipationless chiral currents in experiments. In addition, the phase diagram as functions of Zeeman exchange energy and interlayer distance is also provided in the SM \cite{SM}.


In summary, based on first-principles calculations, we have constructed a WFTB model for a class of ferromagnetic vdWHs composed of 2H-TMDs MX$_2$ (M = Mo or W; X = Se or Te) monolayer and CrY$_{3}$ (Y = Br, I) monolayer. We provide an intuitive picture of valley-splitting  induced by the Zeeman exchange energy of magnetic fields, and the consequent valley-based Hall effects and topological phase transitions have been systematically studied. More importantly, the considerable valley-splitting leads to the VQAHE generating in these vdWHs  with the aid of magnetic proximity effects. Recent experiments have proposed that the magnetism of the 2D ferromagnets can be controlled by gating and external magnetic fields \cite{2018Electrical,2018Controlling}, which can further stimulate the study of tunable topological valley-polarization. Therefore, our work provides a desirable platform utilizing magnetic fields to design topological spin-valley filter for spintronic and valleytronic applications in the future.

~\\~

This work was supported by the National Natural Science Foundation of China (NSFC, Grants No. 11974062, No. 12047564, and No. 11704177), the Chongqing Natural Science Foundation (Grants No. cstc2019jcyj-msxmX0563), the Graduate Scientific Research and Innovation Foundation of Chongqing of China (Grant No. CYS20042), the Fundamental Research Funds for the Central Universities of China (Grant No. 2020CDJQY-A057), and the Beijing National Laboratory for Condensed Matter Physics.



%

\newpage

\begin{widetext}
\newpage

\setcounter{figure}{0}
\setcounter{equation}{0}
\makeatletter

\makeatother
\renewcommand{\thefigure}{S\arabic{figure}}
\renewcommand{\thetable}{S\Roman{table}}
\renewcommand{\theequation}{S\arabic{equation}}

\begin{center}
	\textbf{
		\large{Supplemental Material for}}
	\vspace{0.2cm}
	
	\textbf{
		\large{
			``Magnetic field induced Valley-Polarized Quantum Anomalous Hall Effects in Ferromagnetic van der Waals Heterostructures }
	}
\end{center}

In this Supplemental Material, we give detailed computational methods of first-principles calculations. The electronic band structures, valley splitting, and topological phase transition induced by Zeeman exchange energy of magnetic fields and interlayer distance in other proposed van der Waals heterostructures (vdWHs) are also included.

\section{Computational methods}

The first-principles calculations were performed within the framework of density functional theory (DFT) \cite{PhysRev.136.B864sm,PhysRev.140.A1133sm} using the projector augmented-wave method encoded in the Vienna $ab$ $initio$ Simulation Package (VASP) \cite{PhysRevB.54.11169sm}. The exchange correlation functional was described by the generalized gradient approximation within the Perdew-Burke-Ernzerhof formalism (GGA-PBE) \cite{PhysRevLett.77.3865sm}. The plane-wave cutoff energy was set to be 500 eV. Due to the correlation effects of $3d$ electrons in Cr atoms, we employed the GGA+U method \cite{PhysRevB.52.R5467sm}, in which the effective on-site Coulomb repulsion interaction U was set to 2.0 eV \cite{PhysRevB.101.205404sm}. The first Brillouin zone was sampled by $9 \times 9 \times 1$ Monkhorst-Pack mesh grid. We adopted $1 \times 1$ primitive cell of monolayer chromium trihalides CrY$_{3}$ (Y = Br or I) and $2 \times 2$ supercell of monolayer 2H-phase transition metal dichalcogenides (2H-TMDs) MX$_2$ (M = Mo or W; X = Se or Te) to build heterostructures. A vacuum region larger than 20 \text{\AA} was adopted to decouple the interaction between two neighboring heterostructure slabs. All geometric structures were fully relaxed until energy and force less than $10^{-6}$ eV and 0.01 eV/\text{\AA} using the vdW corrected functional Grimme (DFT-D3) method \cite{Densitysm}. We chose 72 projected atomic orbitals comprised of $d$ orbitals centered at the four W and two Cr sites, $p$ orbitals centered at the eight Te and six I sites in the primitive unit cell to construct Wannier-function-based tight-binding (WFTB) model based on maximally localized Wannier functions methods by the WANNIER90 package \cite{Mostofi2014sm}. The chiral edge states and anomalous Hall conductivity were calculated by the iterative Greens method \cite{Sancho_1985sm} as implemented in the WannierTools \cite{WU2017sm}.

\section{Structures and electrostatic potentials}
We consider three representative stacking configurations between 2H-TMDs and the CrY$_{3}$ (Y = Br or I) substrates in Fig. \ref{Fig. S1}. The vdWHs binding energy is calculated by $E_{\rm b}=E_{\rm vdWHs}-E_{\rm MX_2}-E_{\rm CrY_3}$, where $E_{\rm vdWHs}$, $E_{\rm MX_2}$ and $E_{\rm CrY_3}$  are total energies of vdWHs, monolayer  MX$_2$ and monolayer CrY$_{3}$ respectively. Taking WTe$_{2}$/CrI$_{3}$ vdWH as an example, the Configuration-\uppercase\expandafter{\romannumeral1}, \uppercase\expandafter{\romannumeral2} and \uppercase\expandafter{\romannumeral3} are calculated as $-0.80$ eV, $-0.77$ eV, and $-0.77$ eV, respectively. The most thermodynamically stable structure denotes that one Cr atom in CrY$_{3}$ substrates is located below the Te/Se atom, and another being located below the 2H-TMDs layer hexagon center [Figs. \ref{Fig. S1}(a) and \ref{Fig. S1}(d)]. Table \ref{Tab. S1} lists the parameters of the most stable structure for these vdWHs in detail. All calculations were performed using the most thermodynamically stable structure. In addition, the electrostatic potential of vdWHs are shown in Fig. \ref{Fig. S2}.

\begin{table}[!htbp]
    \caption{Structure properties of vdWHs. The $a$, $d$, $\delta$ and $E_{b}$ represent the optimized lattice constant, averaged layer distance, lattice mismatch and binding energy, respectively.}
  \begin{center}\renewcommand\arraystretch{1.5}
	\begin{tabular}{p{2.5cm}<{\centering} p{2cm}<{\centering} p{2cm}<{\centering} p{2cm}<{\centering} p{2cm}<{\centering}}
		\hline
\hline
		 & $a(\text{\AA})$ & $\delta(\%)$ & $d(\text{\AA})$ & $E_{b}$(eV)\\
       \hline
		WTe$_{2}$/CrI$_{3}$ & 7.01 & 0.8 & 3.71 & -0.80\\
		
		WSe$_{2}$/CrBr$_{3}$ & 6.54 & 2.5 & 3.35 & -0.68\\
		
		MoSe$_{2}$/CrBr$_{3}$ & 6.55 & 2.4 & 3.38 & -0.63\\
		
		MoTe$_{2}$/CrI$_{3}$ & 7.03 & 0.7 & 3.59 & -0.68\\
		\hline
\hline
	\end{tabular}
  \end{center}
     \label{Tab. S1}
\end{table}

\begin{figure*}
    \centering
    \includegraphics[width=12cm]{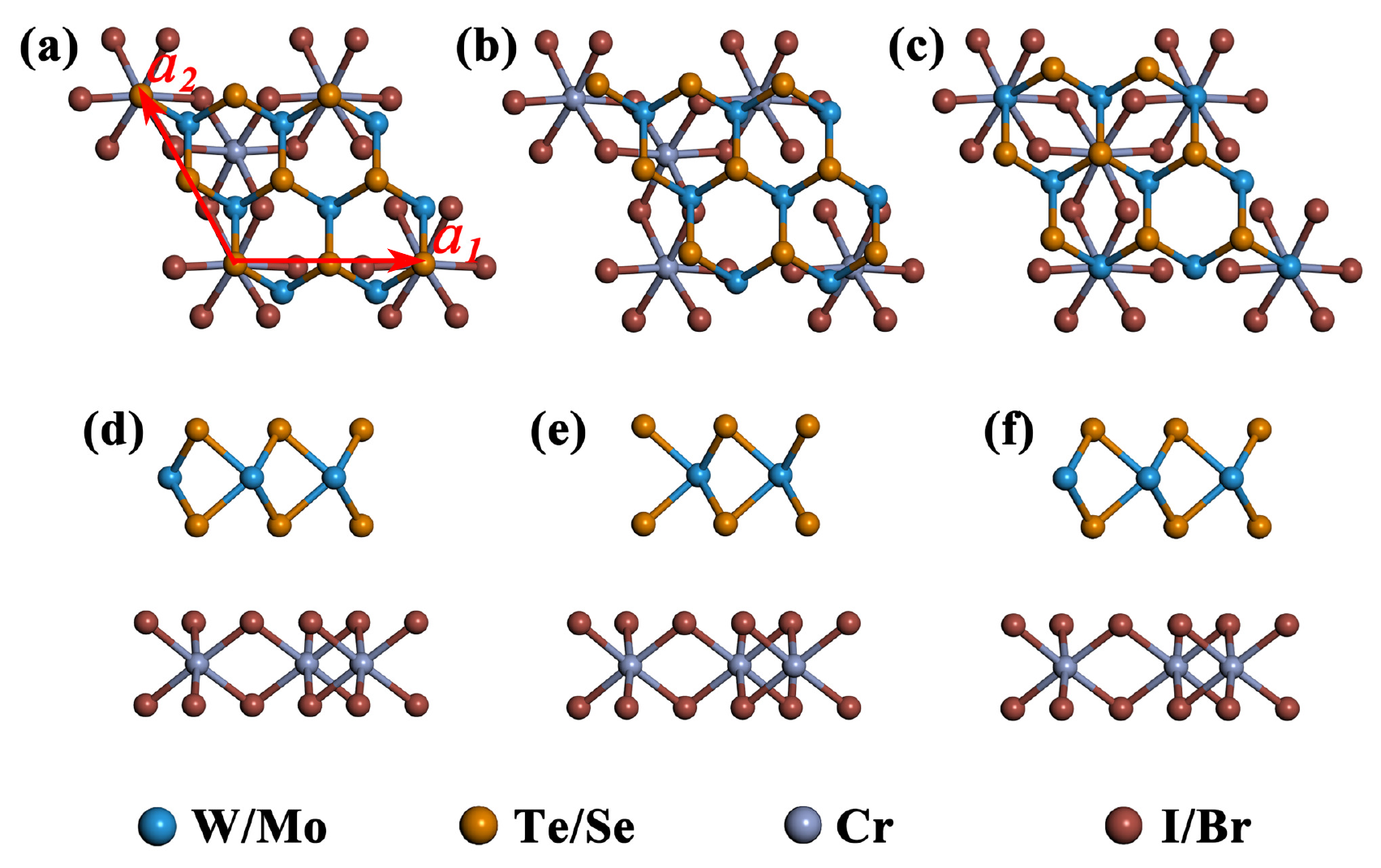}
    \caption{(a)-(c) The top and (d)-(f) side views of three configurations. [(a) and (c)] Configuration-\uppercase\expandafter{\romannumeral1} : one Cr atom is located below of Te/Se atom, another is located below of the TMDs layer hexagon center. [(b) and (e)] Configuration-\uppercase\expandafter{\romannumeral2} and [(c) and (f)] Configuration-\uppercase\expandafter{\romannumeral3} are obtained respectively by moving the TMDs layer along the $\frac{1}{12}(2\mathbf{a}_{1}+\mathbf{a}_{2})$ and $\frac{1}{6}(2\mathbf{a}_{1}+\mathbf{a}_{2})$ with respect to the CrY$_{3}$ layer.}
    \label{Fig. S1}
\end{figure*}

\begin{figure*}
    \centering
    \includegraphics[width=17.3cm]{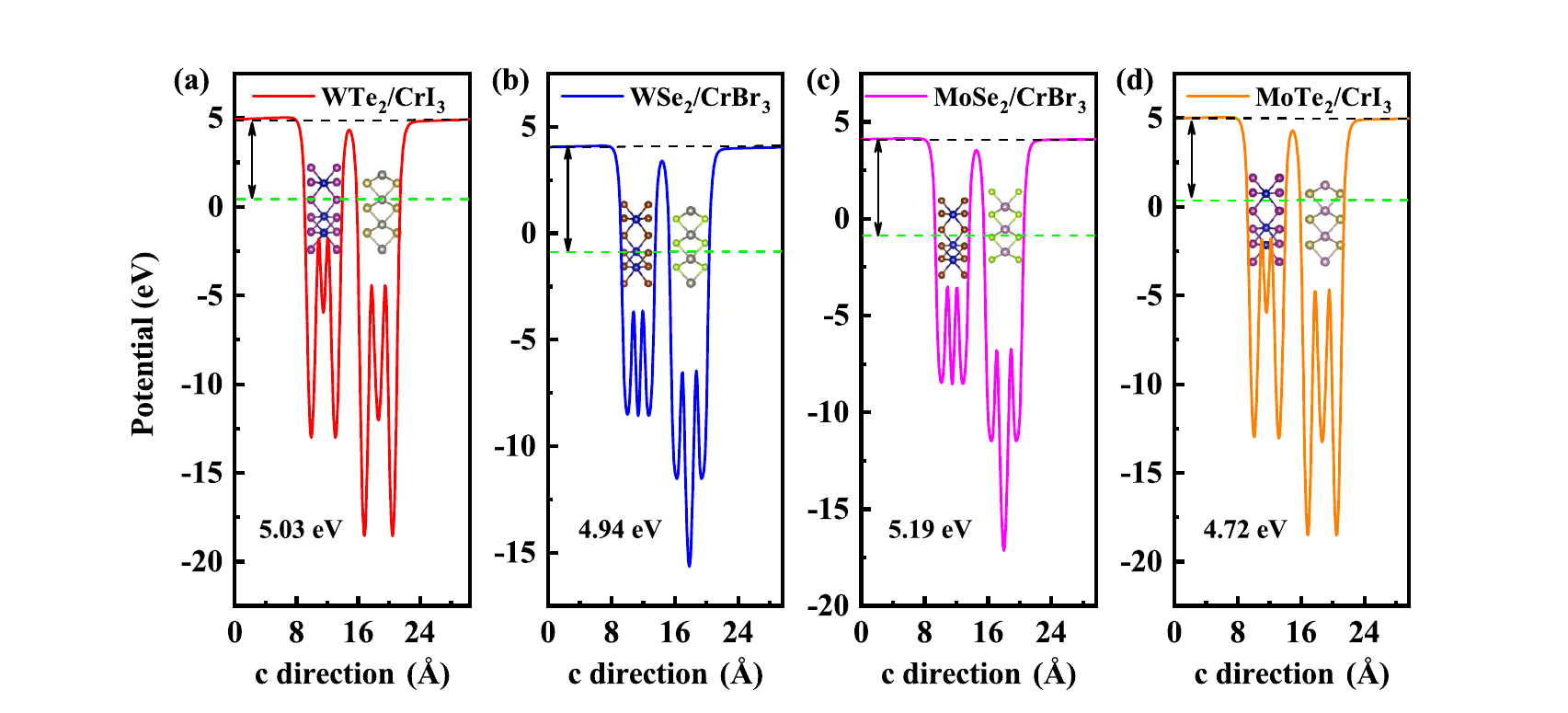}
    \caption{The electrostatic potential of (a) WTe$_{2}$/CrI$_{3}$, (b) WSe$_{2}$/CrBr$_{3}$, (c) MoSe$_{2}$/CrBr$_{3}$ and MoTe$_{2}$/CrI$_{3}$. The black and green dotted line represents respectively the vacuum level ($E_{vac}$) and Fermi level ($E_{f}$).}
    \label{Fig. S2}
\end{figure*}

\section{Valley splitting under magnetic field and electronic band structures of the other vdWHs}

\begin{figure*}
    \centering
    \includegraphics[width=17.3cm]{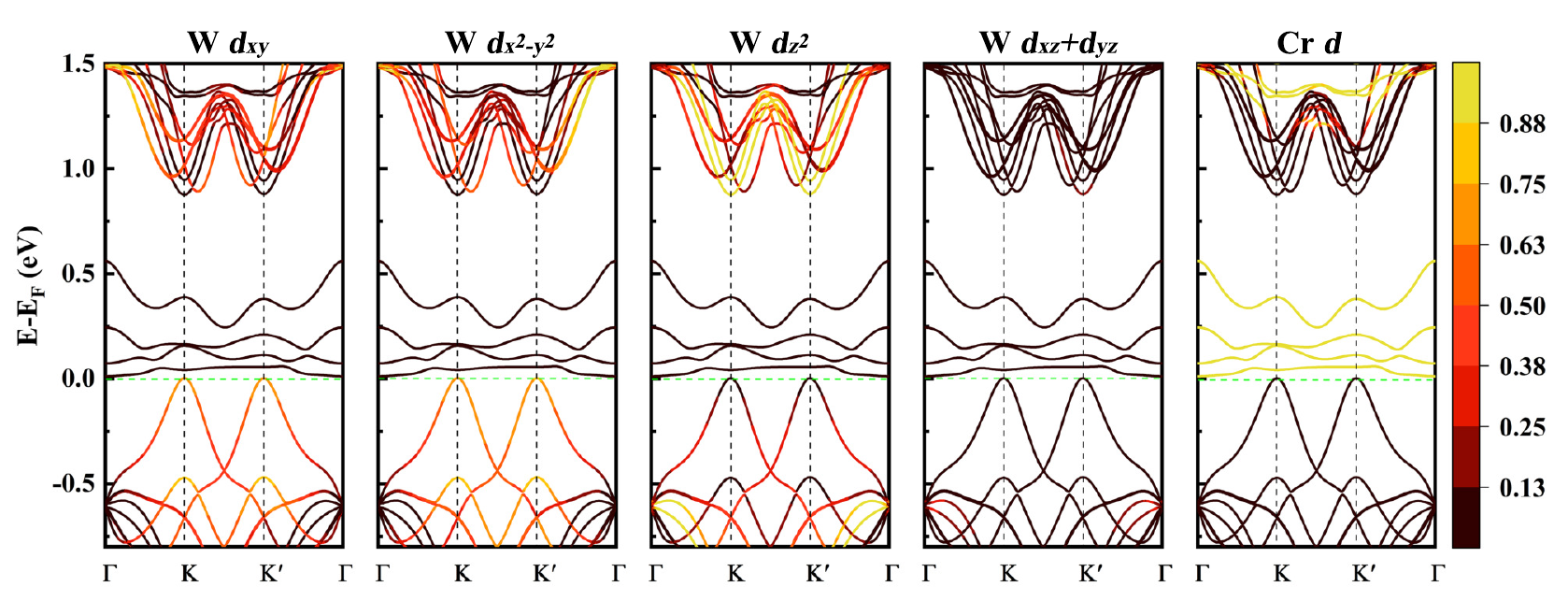}
    \caption{The orbital components in the electronic band structures of WTe$_{2}$/CrI$_{3}$ vdWH.}
    \label{Fig. S3}
\end{figure*}
\begin{figure*}
    \centering
    \includegraphics[width=12cm]{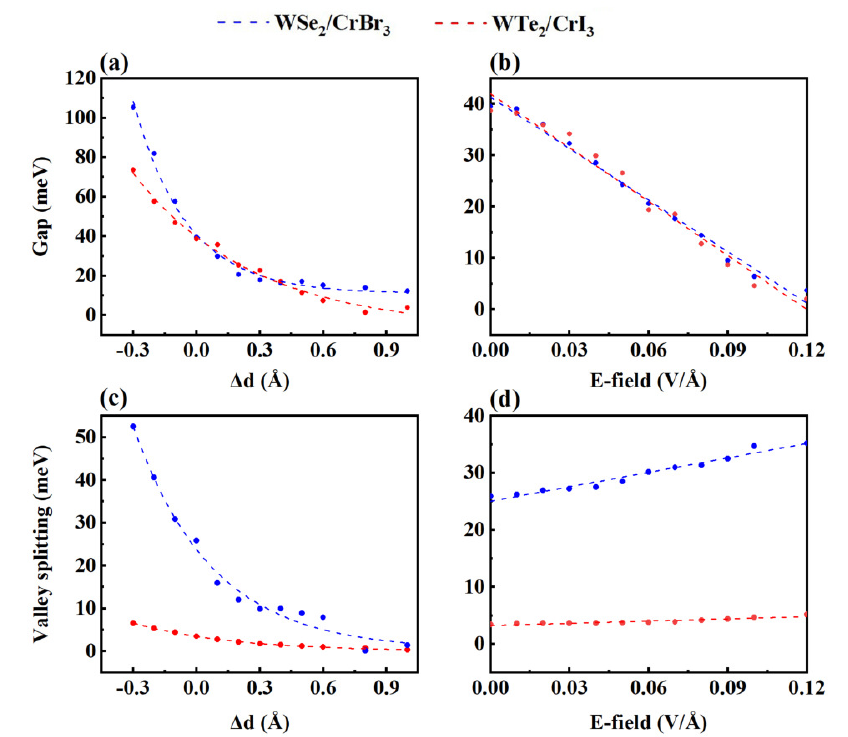}
    \caption{Upon consideration of SOC, [(a) and (b)] the band gap at the $K$ point (or [(c) and (d)] valley splitting) as a function of the interlayer distance $\varDelta\textit{d}$ with respect to equilibrium position or external electric field in WTe$_{2}$/CrI$_{3}$ and WSe$_{2}$/CrBr$_{3}$ vdWH.}
    \label{Fig. S4}
\end{figure*}

The valley splitting can be understood with the interaction between magnetic field and atomic orbital contribution \cite{2014Magneticsm,PhysRevLett.114.037401sm}. As illustrated in Fig. \ref{Fig. S3}, the conduction band edges of WTe$_{2}$ are predominantly composed of $d_{z^{2}}$ orbital of W atoms with magnetic quantum number $m=0$, whereas the valence band edges are predominantly $d_{xy}$ and $d_{x^{2}-{y^{2}}}$ orbitals of W atoms with $m=\pm2$ in the $K$/$K^\prime$ valley. Thus, this contribution will not shift to the conduction band edges and a shift of $\varDelta=2\tau\mu_BB$ to the valence band edges, where $\tau=\pm1$ is the index for the  $K$/$K^\prime$ valley. On the whole, the contribution to valley splitting is $\varDelta^{K}-\varDelta^{K^\prime}=-4\mu_BB$  from atomic orbit magnetic moments.

Magnetic proximity interaction modulated by the interlayer distance $\Delta d$ and external electric ($\mathbf{E}$) field as shown in Fig. \ref{Fig. S4},  It is found that the valley splitting increases exponentially with the decrease of $\Delta d$, as shown in Fig. \ref{Fig. S4} (d), corrspongding to the fitted exponential function $Ae^{-\Delta d/\gamma}$, where $A$ and $\gamma$ are the fitting parameters (the fitting parameters $A$=3.55, $\gamma$=0.47 for WSe$_{2}$/CrBr$_{3}$ and $A$=23.51, $\gamma$=0.38 for WTe$_{2}$/CrI$_{3}$). In addition, the valley splitting increase linearly with the external $\mathbf{E}$ field, corrspongding to the fitted linear function $x\mathbf{E}$ with slope $x=84.31$ for WSe$_{2}$/CrBr$_{3}$ and $x=13.01$ for WTe$_{2}$/CrI$_{3}$ vdWH.

\section{Topological phase transition in WTe$_{2}$/CrI$_{3}$ vdWH}

\begin{figure*}
    \centering
    \includegraphics[width=12cm]{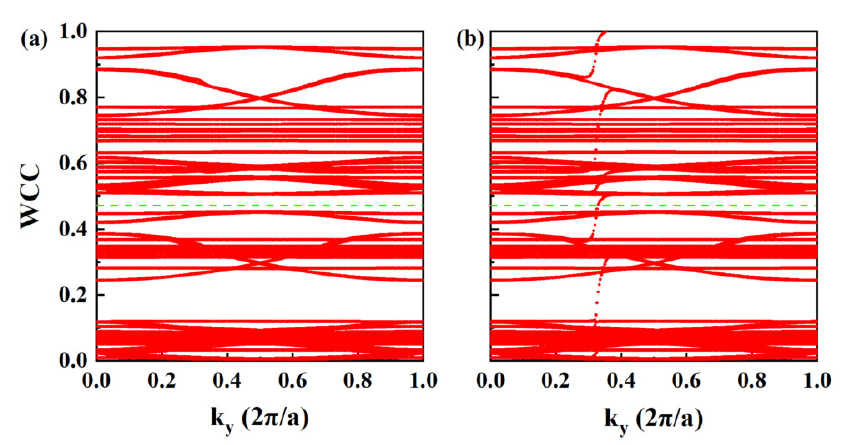}
    \caption{(a) Evolution of Wannier charge centers (WCCs) as a function of $k_{y}$ (a) without external magnetic field and (b) with external magnetic field exchange energy 35 meV, when a perpendicular $\mathbf{E}$ field is fixed at 0.1 (V/\AA) in WSe$_{2}$/CrBr$_{3}$ vdWH.}
    \label{Fig. S5}
\end{figure*}

\begin{figure*}
    \centering
    \includegraphics[width=13cm]{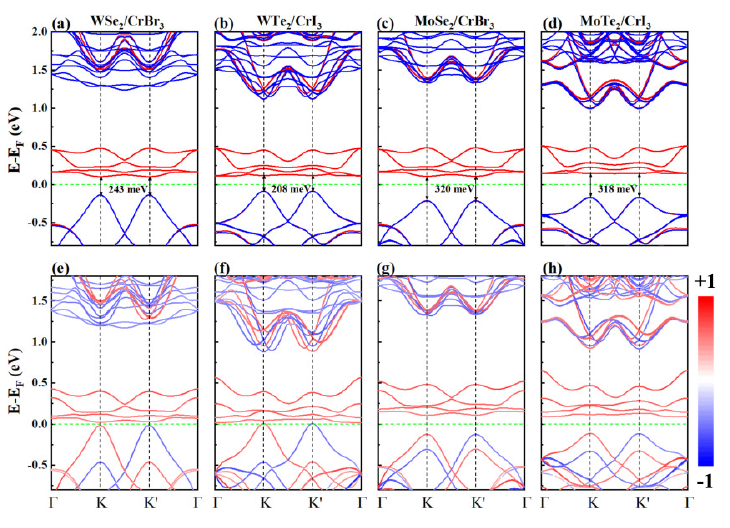}
   \caption{Band structures of WTe$_{2}$/CrI$_{3}$, WSe$_{2}$/CrBr$_{3}$, MoSe$_{2}$/CrBr$_{3}$ and MoTe$_{2}$/CrI$_{3}$ heterostructures without [(a)-(d)] and with [(e)-(h)] SOC. Red (blue) lines denote spin-up (spin-down) channels.}
    \label{Fig. S6}
\end{figure*}

As shown in Fig. \ref{Fig. S6}, the similar band structures with small energy gap at the $K$/$K^\prime$ valley are found in WTe$_{2}$/CrI$_{3}$, WSe$_{2}$/CrBr$_{3}$, MoSe$_{2}$/CrBr$_{3}$ and MoTe$_{2}$/CrI$_{3}$ heterostructures. The detailed topological phase transitions of WTe$_{2}$/CrI$_{3}$ are illustrated in Figs. \ref{Fig. S7} and \ref{Fig. S8}. Our results demonstrate that it is general strategy that utilizing magnetic field realizes valley polarized topological phase transitions in these 2D ferromagnetic heterostructures.

\begin{figure*}
    \centering
    \includegraphics[width=17.3cm]{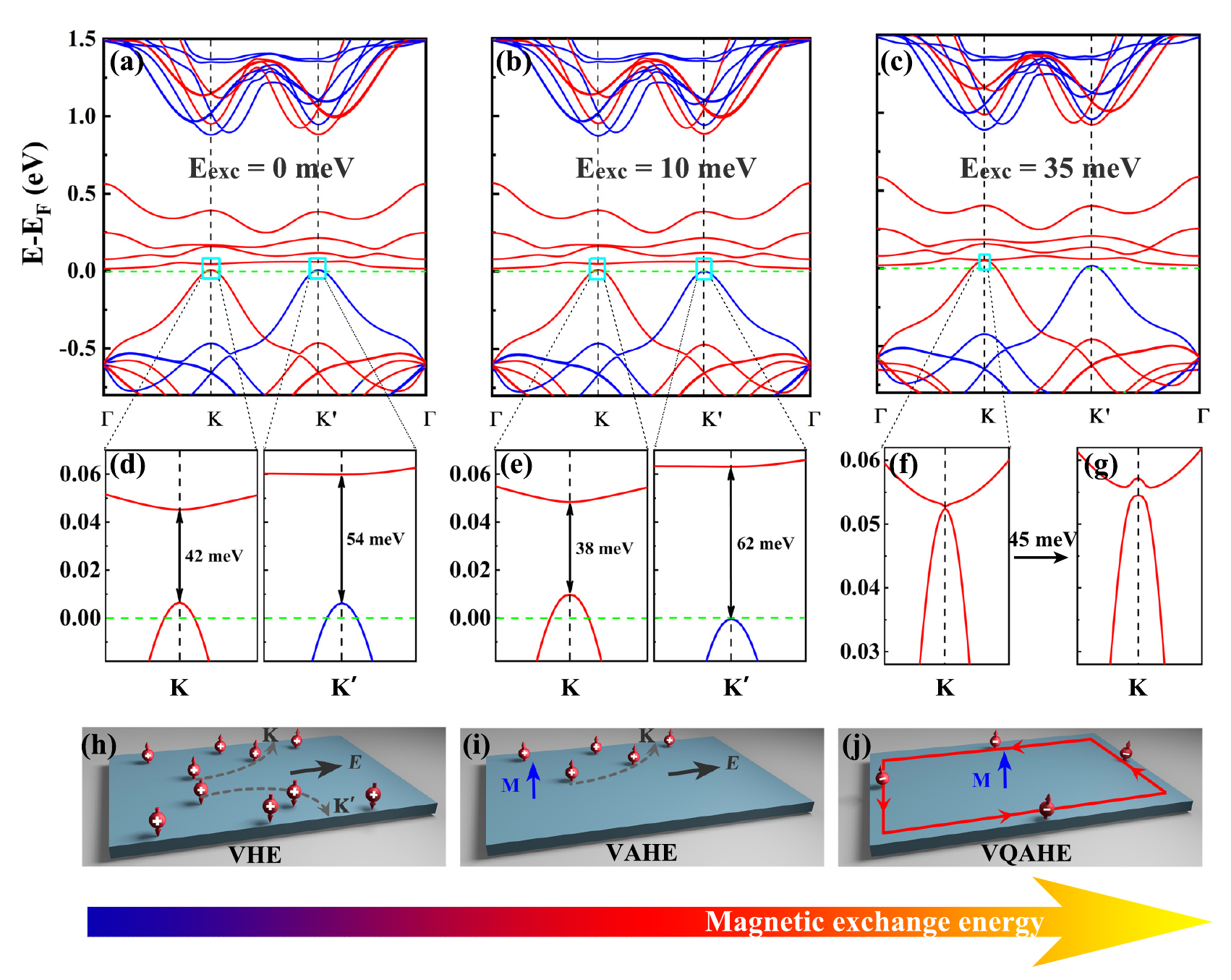}
    \caption{(a)-(c) In WTe$_{2}$/CrI$_{3}$ vdWH, upon consideration of SOC, the spin-resolved band structure evolution as the external magnetic field exchange energy, when interlayer distance is expanded 0.5 \AA. (d)-(f) The corresponding enlarged views at the $K$ and $K^\prime$ valleys. Red (blue) lines denote spin-up (spin-down) channels. (g) Band inversion at the $K$ valley when the exchange energy further increased 45 meV. (h) and (j) Schematic diagram of topological phase transition from VHE to VQAHE with increasing the Zeeman exchange energy of magnetic field. }
    \label{Fig. S7}
\end{figure*}

\begin{figure*}
    \centering
    \includegraphics[width=12cm]{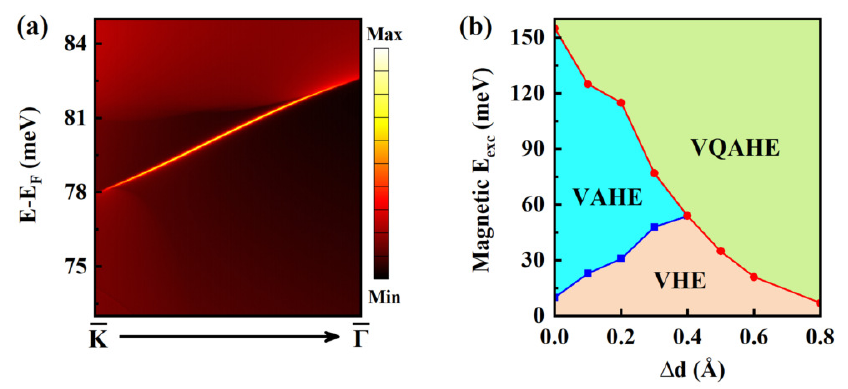}
   \caption{(a) The calculated edge states of the semi-infinite  WTe$_{2}$/CrI$_{3}$. (b) The valley polarized Hall effect phase diagram for WTe$_{2}$/CrI$_{3}$ vdWH in terms of the interlayer distance $\varDelta\textit{d}$ and Zeeman exchange energy.}
    \label{Fig. S8}
\end{figure*}

\vspace{100em}
~~~\\

%

\end{widetext}
\end{document}